# Electronic Structure of $C_{60}$/Graphite

P. A. Brühwiler, P. Baltzer, S. Andersson, D. Arvanitis, and N. Mårtensson

*Dept. of Physics, Uppsala University, Box 530, SE-751 21 Uppsala, Sweden*

**Abstract.** We report temperature-dependent photoelectron spectra for a monolayer of $C_{60}$ adsorbed on HOPG, as well as C $1s$ x-ray absorption. This extends a previous report which showed the close similarity between the spectrum of the HOMO for the two-dimensional overlayer and that of $C_{60}$ in the gas phase. The present work shows that intermolecular and molecule-substrate vibrations contribute strongly to the spectral lineshape at room temperature. Thus, vibrational effects are shown to be crucial for the proper understanding of photoelectron spectra, and thus the charge transport properties, for $C_{60}$ in contact with graphite and graphite-like materials.

## INTRODUCTION

The interaction of $C_{60}$ and graphite is fundamentally interesting, comprising as it does the pairing of two basic, van der Waals-bound carbon materials. It has been studied by several groups, both theoretically [1, 2, 3, 4, 5, 6] and experimentally [7, 8, 9, 10, 11, 12, 13]. The recent discovery and characterization of $C_{60}$ and other fullerenes contained within carbon nanotubes [14, 15], so-called "peapods", has extended the interest in this interaction into new areas, including the construction of devices based on encapsulated fullerenes [16, 17]. Here we present Photoelectron (PES) and X-ray (XAS) spectroscopic data which give some insight into the electronic structure and the role of vibrations for lower-dimensional $C_{60}$ structures in contact with graphite.

## EXPERIMENTAL

Sample preparation has been described in some detail previously [10]. Here we would like to note that the film was deposited at 110 K, and the monolayer morphology was optimized by warming the sample to 450 K for one minute. This was found to narrow the C 1s line to 0.38 eV from 0.50 eV, simultaneously reducing the binding energy from 285.00 to 284.93 eV. The resulting width is quite close to the value for a single layer van der Waals bound to a $C_{60}$ monolayer (see, e.g., [18]), and the shift in energy is consistent with the elimination of the $C_{60}$ surface core level shift of 0.1 eV [19]. At a coverage of 0.2 ML for an annealed overlayer, one can surmise that large islands form [7, 11] so that the vast majority of the $C_{60}$ molecules probed in the experiment have six nearest neighbors as in a perfect monolayer [10]. Thus our assignment of the coverage to a single-layered film is also in agreement with calculations showing that island edges should stabilize single layers over double layers [6].

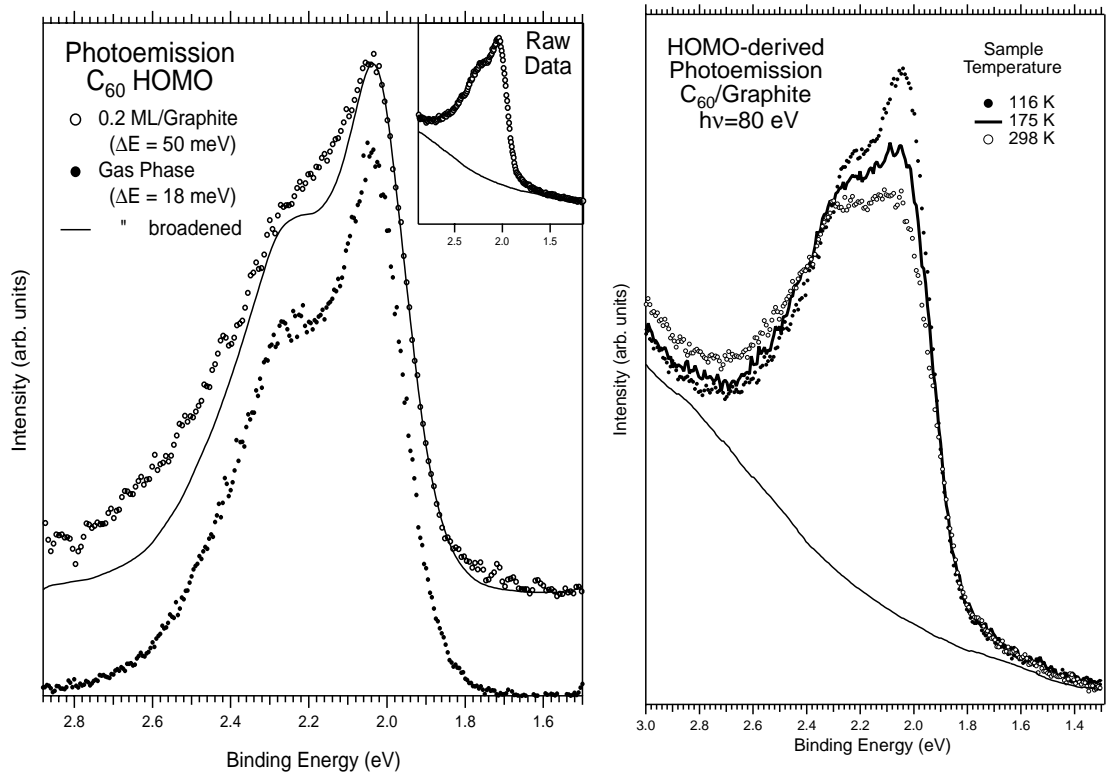

**FIGURE 1.** (a) PES of the indicated samples, measured at normal emission. The solid line is the gas phase spectrum after broadening to correspond to the same instrumental resolution as the overlayer data. Inset: PES data for the overlayer, before subtracting the background which is a spectrum taken on the pristine HOPG surface (the latter, also shown, was smoothed slightly before subtraction). At normal emission, the background is relatively featureless near $E_F$, as is seen from the spectrum. (b) PES spectra of the HOMO-derived band at the indicate temperatures, showing the effects of intermolecular and $C_{60}$-graphite vibrations.

## RESULTS AND DISCUSSION

The photoemission data are presented in Fig. 1(a). Since the gas phase spectrum is broadened purely by Jahn-Teller vibrational coupling in the final state [10], the strong similarity between the gas phase and monolayer data suggests that any two-dimensional band structure effects are well-masked by the intramolecular vibronic coupling [10]. The expected size of the electronic bandwidth in two dimensions rests largely on the correct determination of the intermolecular spacing. An early determination using a sensitive form of Low Energy Electron Diffraction (LEED) found a spacing of $10.1 \pm 0.1$ Å, quite close to that of solid $C_{60}$ [9], whereas a more recent study using the same technique and with similar k-resolution derived a spacing closer to 10.5 Å [12]. Hence, the question of the monolayer intermolecular distance, and the expected magnitude of polaronic effects in determining the effective mass of (hole) carriers in condensed $C_{60}$ [10], awaits a resolution of the conflict between these studies. In any case, the $C_{60}$/graphite interaction is seen to be quite weak, consistent with van der Waals bonding.

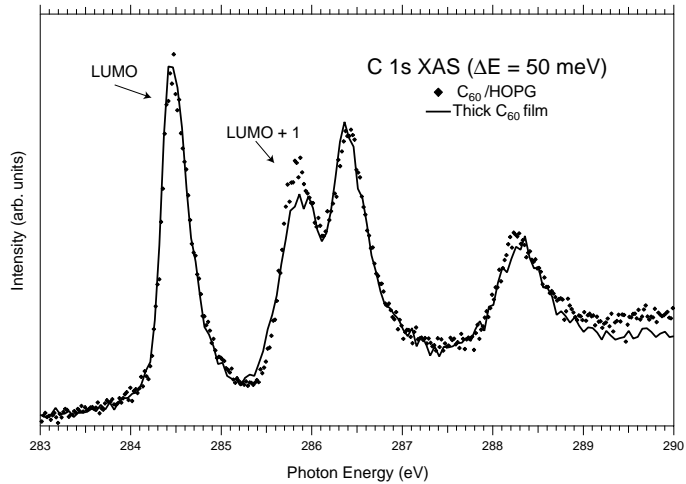

**FIGURE 2.** C 1$s$ XAS of the indicated samples. The data for 0.2 ML were taken at a temperature of 116 K, and with the light at normal incidence, so that the graphite signal in the indicated energy range is very low and structureless (becomes noticeable at around 289 eV, the onset of the graphite $\sigma^*$ signal). The data shown are the highest resolution spectra we measured (of the order of 50 meV photon bandpass).

To further elucidate the role of vibrations, we present temperature-dependent data in Fig. 1(b). Here we see that the spectra are quite sensitive to temperature, illustrating the important role of intermolecular vibrations and molecule/substrate vibrations in the lineshape observed at room temperature. A crude estimate of the broadening expected can be obtained from theoretical van der Waals intermolecular potentials, and suggests a contribution of about 35 meV per nearest neighbor $C_{60}$ molecule, as well as a similar contribution from the graphite substrate. We note that lattice expansion is a possible consequence of the increasing temperature as well. Why such temperature-dependence is not observed more generally for spectra from solid $C_{60}$ films remains an open question.

We present C 1$s$ absorption data for this system in Fig. 2 and compare to data of solid $C_{60}$. Quite apparent is the general similarity of the two spectra, as expected for a neutral excitation in van der Waals-bonded system. This is emphasized by the virtually identical LUMO-resonance width–the LUMO resonance can be shown to lie in the fundamental bandgap of solid $C_{60}$ [20, 21, 22], and thus is broadened entirely by intramolecular vibrations. The strong correspondence of the two LUMO resonances suggests that intramolecular vibrations are the dominant broadening factor for the monolayer case as well. From the graphite C 1$s$ binding energy of 284.4 eV, which marks $E_F$ for the spectra [23], it is clear that all the resonances for the monolayer overlap the continuum of $\pi^*$-bands of the substrate, and might therefore exhibit broadening due to a bonding interaction. The similarity of the spectra thus shows that any such bonding interaction is quite weak. It is perhaps worth noting that a similar narrowing of the second resonance has also been observed for a layer of $C_{60}$ adsorbed on a monolayer of $C_{60}$/Au(110) [18], as well as for matrix-isolated $C_{60}$ [20]. Thus the second LUMO resonance is a sensitive marker of bonding effects. Furthermore, corresponding data for $C_{60}$ inside carbon nanotubes appear to be characterized by the same narrowing of the second resonance, showing a strong similarity to the bonding to graphite for that system [24].

In summary, we have taken up the issue of the bonding of $C_{60}$ to graphite from an experimental viewpoint, and shown that this bonding has small (PES) or negligible (XAS) impact on a purely molecular picture of the spectra, consistent with van der Waals interactions. Intramolecular vibrations are quite important for explaining the width of the features observed for isolated molecules, and for the two-dimensional overlayer studied here. One must also account for significant contributions from intermolecular vibrations, especially at room temperature. These effects should significantly modify electron scattering via the fullerene levels for hot electron transport in "peapods" recently discussed [17]. It remains to be seen if a quantitative study of these effects for solid $C_{60}$ can isolate the role of vibrations and vibronic coupling for the electronic states of this molecular solid.

## ACKNOWLEDGMENTS

We would like to thank A. J. Maxwell for help with the measurements, and the staff of MAX-Lab for general assistance. We would also like to acknowledge the Consortium on Clusters and Ultrafine Particles for financial support, which in turn is supported by Stiftelsen för Strategisk Forskning, as well as Naturforskningsrådet.